\documentclass[review]{elsarticle}

\usepackage{lineno,hyperref}
\usepackage{graphicx}
\modulolinenumbers[5]

\journal{JAC}

\begin{document}


\title{Effects of proton irradiation on flux-pinning properties of underdoped Ba(Fe$_{0.96}$Co$_{0.04}$)$_2$As$_2$ pnictide superconductor}


\author{S. Salem-Sugui Jr.$^1$, D. Moseley$^2$, S. J. Stuard$^1$, A. D. Alvarenga$^3$, A. S. Sefat$^4$, L. F. Cohen$^2$, L. Ghivelder$^1$}

\address{$^1$Instituto de Fisica, Universidade Federal do Rio de Janeiro,
 Rio de Janeiro, RJ, 21941-972, Brazil}

\address{$^2$Blackett Laboratory, Imperial College London, London, SW7 2AZ, UK}

\address{$^3$Instituto Nacional de Metrologia Qualidade e Tecnologia, Duque de Caxias, RJ, 25250-020, Brazil}

\address{$^4$Oak Ridge National Laboratory, Oak Ridge, TN 37831, USA}

\begin{abstract}
We study the effect of proton irradiation on Ba(Fe$_{0.96}$Co$_{0.04}$)$_2$As$_2$ superconducting single crystals from combined magnetisation and magnetoresistivity measurements. The study allows the extraction of the values of the apparent pinning energy $U_0$ of the samples prior to and after irradiation, as well as comparison of the values of $U_0$ obtained from the flux-flow reversible region with those from the flux-creep irreversible region. Irradiation reduces $T_c$ modestly, but significantly reduces $U_0$ in both regimes: the critical current density $J_c$ is modified, most strikingly by the disappearance of the second magnetisation peak after irradiation. Analysis of the functional form of the pinning force and of the temperature dependence of $J_c$ for zero field, indicates that proton irradiation in this case has not changed the pinning regime, but has introduced a high density of shallow point-like defects. By consideration of a model that takes into account the effect of disorder on the irreversibility line, the data suggests that irradiation produced a considerable reduction in the average effective disorder overall, consistent with the changes observed in $U_0$ and $J_c$. 
\end{abstract}
\begin{keyword}
\texttt{pnictides, proton irradiation, superconducting critical current, flux-pinning}

\end{keyword}



\maketitle

\section{Introduction}
Since the discovery of the iron pnictides superconductors \cite{refpin}, flux-pinning has been widely studied in most of the iron-superconducting systems, in order to evaluate the potential of this material for application \cite{refjo,beek,beek2,yamamoto}. As previously studied in copper based high-$T_c$ superconductors \cite{david,ger,kon,sauer}, the study of flux-pinning after proton and particle irradiation in pnictides has gained increasing interest \cite{rev,1,2,3,4,5,BaK,phyc,lesley,FeCo}. Irradiation allows the effect of disorder on the vortex state to be studied by changing the flux-pinning, leading in some cases to the disapearance of the second magnetisation peak, SMP, (often described as the fishtail peak) \cite{rev}. Depending on the irradiation dose, it has been shown that the effect of disorder can reduce the value of $T_c$ and potentially also change the Fermi level in the bands participating in the pairing mechanism.\cite{6}

In this work, we study the effect of proton irradiation on underdoped Ba(Fe$_{0.96}$Co$_{0.04}$)$_2$As$_2$ single crystals \cite{sample,lester}. For Ba(Fe$_{1-x}$Co$_{x}$)$_2$As$_2$ system, the underdoped content of Co occur for $x$$<$0.06 \cite{lester}. The investigation presents a comparative study performed on two crystals from the same batch, with the same superconducting transition temperatures, $T_c$ = 13 K, after which one of the crystals was irradiated, producing a drop in the value of the transition temperature, to $T_c$ $\sim$ 12 K and the disappearance of the SMP. The study was conducted by obtaining isothermal resistivity curves as a function of magnetic field and isothermal magnetisation curves as a function of magnetic field and time (isofield magnetic relaxation curves) on both crystals. Results of the analysis allowed the comparison of the effect of irradiation on the activation energy, $U_0$, as well as the comparison for each sample, of the values of $U_0$ obtained from resistivity taken in the reversible regime, with the $U_0$ values obtained from flux-creep, in the irreversible regime. We also study the effect of irradiation on the normalized volume pinning force, on the temperature dependent behavior of the normalized remanent critical current, and on the irreversible lines which were fitted with an expression\cite{baruch} that allows the estimation of the disorder within each sample.
\section{Experimental}
The high-quality Ba(Fe$_{0.96}$Co$_{0.04}$)$_2$As$_2$ single crystals used in this work with approximate dimensions: crystal 1 (virgin)  2.7$\times$2.2$\times$0.19 mm$^3$ and mass $m$ = 5.72 mg, crystal 2 (irradiated) 0.9 $\times$0.8$\times$0.025 mm$^3$ and mass $m$ = 0.85 mg and crystal 3 (irradiated) 0.9$\times$0.75$\times$0.02 mm$^3$ and mass 0.96 mg, were grown by the CoAs flux method \cite{sample}. Crystal 3 was only studied by electrical transport and was damaged after a unique isothermal $M$($H$) at $T$ = 6 K was obtained. The crystals prior to irradiation exhibit a  $T_c$ = 13 K with $\delta$$T_c$ $<$ 1 K. Proton irradiation has been shown to introduce defects,
which are recognized to be predominantly of a point defect character, without significantly altering the intrinsic electronic structure of the material \cite{lc1,lc2}. Proton irradiation induces charge carrier scattering, increasing the residual electrical resistivity $\delta$$\rho$(0), and lowering both $T_N$, the Neel temperature transition, and $T_c$ \cite{lc1,lc3}.  In the present study proton irradiation was performed such that a proton dose of 0.5x10$^{16}$ cm$^{-2}$ (crystal 3) and 1.5x10$^{16}$ cm$^{-2}$ (crystal 2) were achieved for the Ba(Fe$_{0.96}$Co$_{0.04}$)$_2$As$_2$ crystals. Crystal sizes were sufficiently thin ($<$50 $\mu$m) to achieve homogeneous irradiation damage using 3-MeV H+ irradiation according to calculations using the SRIM code. The observed ratio of ($\delta$$T_c$/$\delta$$\rho$(0)) with irradation, produced a value of -0.02 K($\mu$$\Omega$cm)$^{-1}$ which is significantly lower than the -0.08 K($\mu$$\Omega$cm)$^{-1}$  calculated by Nakajima et al \cite{lc1} on Ba(Fe$_{0.955}$Co$_{0.045}$)$_2$As$_{2}$. In the present work, the irradiation was conducted at room temperature while in comparable studies low temperature proton irradiation (50 K) was utilized. Taen et al \cite{lc4} have suggested that the random point defects generated by the irradiation may be annealed out of the structure at room temperature. This possibility is unlikely to have occurred in the present study as the samples were properly thermally anchored by mounting on a large copper block during irradiation avoiding any possible overheating. This is corroborated by the increase of residual resistivity and decrease of $T_c$ and $T_N$ (as shown below when discussing Fig. 1) indicating that additional defects have been consistently created by the irradiation. 

All transport measurements were taken using the van der Pauw technique \cite{lc5} with the magnetic field applied along the c axis and current within the ab plane.
Magnetisation measurements as a function of temperature, $M(T)$, and applied field, $M(H)$, were made with a Quantum Design VSM-9T system. The results were obtained after cooling the crystals from above $T_c$ to the desired temperature in zero applied magnetic field (ZFC), after which the magnetic field is applied  along the c-axis. Magnetic hysteresis loops were obtained for various temperatures below $T_c$ in each crystal, and magnetic relaxation curves were obtained for selected magnetic fields and temperatures over span times extending up to 3 hours. 


\section{Results and discussion}

\begin{figure}[t]
\includegraphics[width=\linewidth]{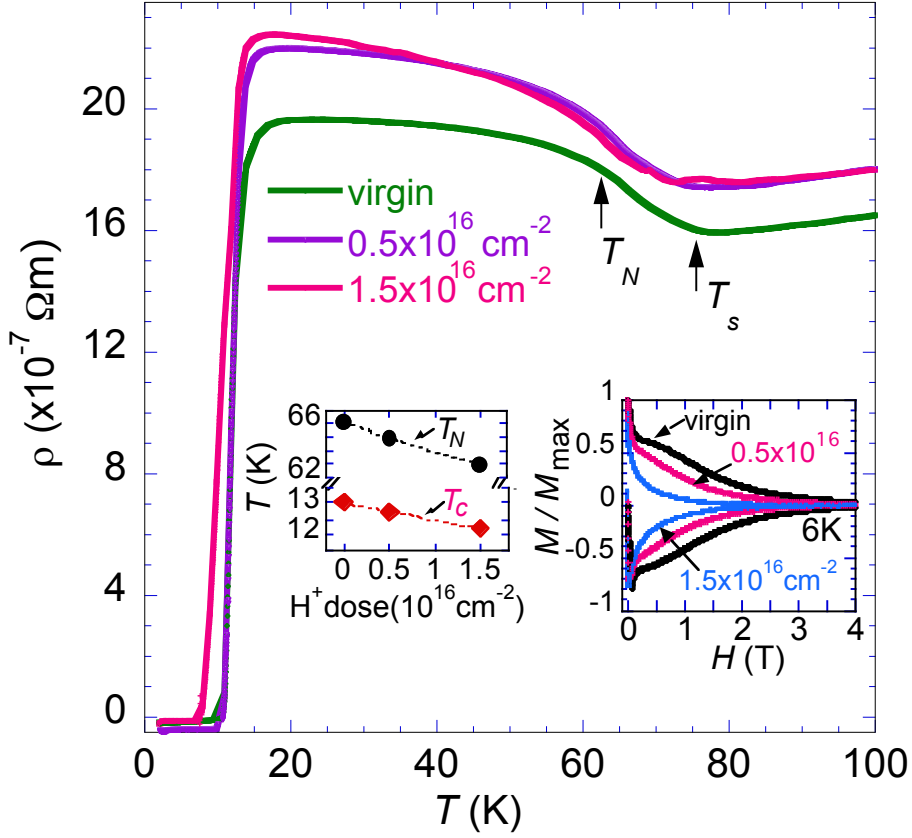}
\caption{Resistivity curves for three different crystals after different doses of proton irradiation. Insets: $T_c$ and $T_N$ as a function of irradiation doses; isothermal $M$($H$) curves at $T$ = 6 K as a function of irradiation doses.}
\label{fig1}
\end{figure}

\begin{figure}[t]
\includegraphics[width=\linewidth]{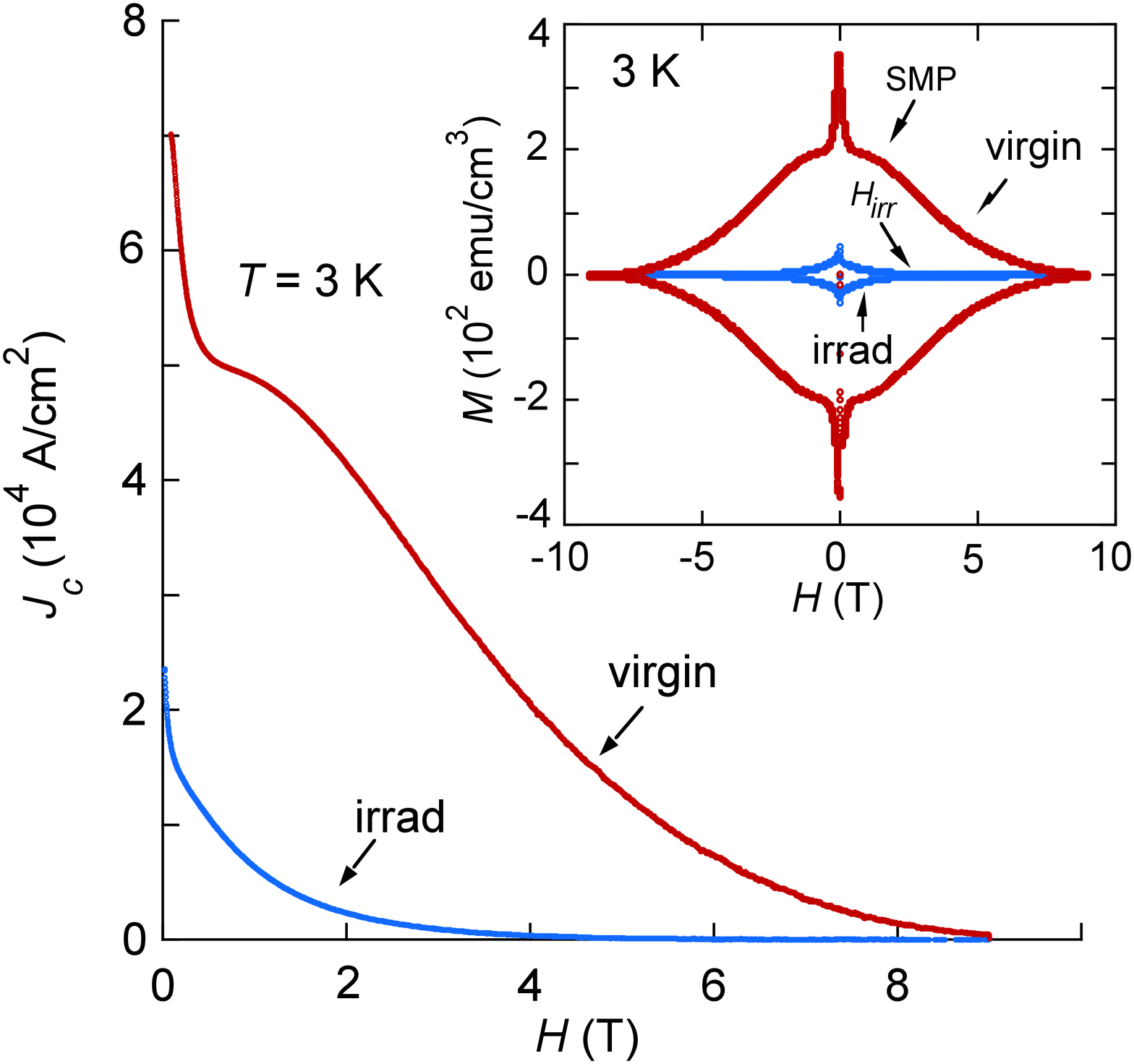}
\caption{Isothermal $J_c$($H$) curves as a function of magnetic field at $T$ = 3K. inset: isothermal magnetisation versus field hysteresis curves as obtained for crystal 1 and 2 at $T$ = 3K.}
\label{fig2}
\end{figure}

Figure 1 shows a plot of resistivity as a function of temperature for an unirradiated crystal and two other irradiated crystals after exposed to different doses. As already mentioned, both $T_N$ and $T_c$ are lowered after irradiation, which is shown in an inner inset of Fig. 1. The values of $T_c$ in this inset were obtained by the 80 $\%$ value of the normal resistivity criterion. A second inset in Fig. 1 shows the effect of irradiation on isothermal $M$($H$) curves obtained at $T$ = 6 K. The data on crystal 3 is included only to illustrate the effect of irradiation on $T_c$ on $T_N$ and on the $M$($H$) curves. Figure 1 and the insets show that the low dose of proton irradiation reduced both the value of $T_c$ and of the transition width $\delta$$T_c$/$T_c$, as observed in optimally doped Ba(Fe$_{0.93}$Co$_{0.07}$)$_2$As$_2$ \cite{FeCo} ($T_c$ = 24 K), but produced an overall decrease of the magnetization values when compared with the unirradiated sample. The highest dose of irradiation reduced even further the value of $T_c$ but increased the transition width. This apparent unusual behavior was also observed in underdoped samples of (Ba-K)Fe$_2$As$_2$ after high doses of proton irradiation \cite{BaK}. The highest dose of irradiation also produced an additional, consistent, decreasie of the values of magnetization, as shown in the right inner inset.

Figure 2 shows the effect of irradiation on the estimated critical current density at $T$ = 3K. The critical current density is given by \cite{7} $J_c$ = 20$\Delta$$M$/($a$(1-$a$/3$b$)) , where $\Delta M$ is the full width of each hysteresis curve shown in the upper inset of Fig. 2 and $b$$>$$a$ are the dimensions of the samples. Arrows in the inset figure marks the region where the second magnetisation peak, SMP, develops in the virgin sample and shows the irreversible field, $H_{irr}$, for the irradiated sample. As shown in Fig. 2, irradiation produces a large reduction of $J_c$ and the disappearance of the SMP. 

To further study the effect of irradiation on flux pinning we perform magnetic relaxation measurements ($M$ vs. time) on both crystals at fixed temperatures and fixed applied magnetic fields. Figure 3a shows selected isofield ln$M$ vs ln$t$ curves obtained at $T$ = 3 K for the virgin sample. Logarithmic behavior with time was observed in all relaxation curves, on both crystals, allowing the apparent activation energy $U_0$ defined by dln$M$/dln$t$=$k_B$$T$/$U_0$, to be obtained. Figure 3b shows isofield plots of the logarithm of the resistivity as a function of 1/$T$ as extracted from isothermal magnetoresistivity curves obtained for the irradiated sample. Similar plots were obtained for the virgin sample following the same procedure.  As shown in the curves of Fig. 3b, an approximately linear behavior of ln$R$ occurs in the lower region of the curves, which appears to be associated with the well known thermally assisted flux flow resistivity \cite{PKess}, where ln$R$ = ln$R_0$-$U_0$/$k_B$$T$, allowing the flux-flow activation pinning energy $U_0$ to be estimated for each curve.

 \begin{figure}[t]
\includegraphics[width=\linewidth]{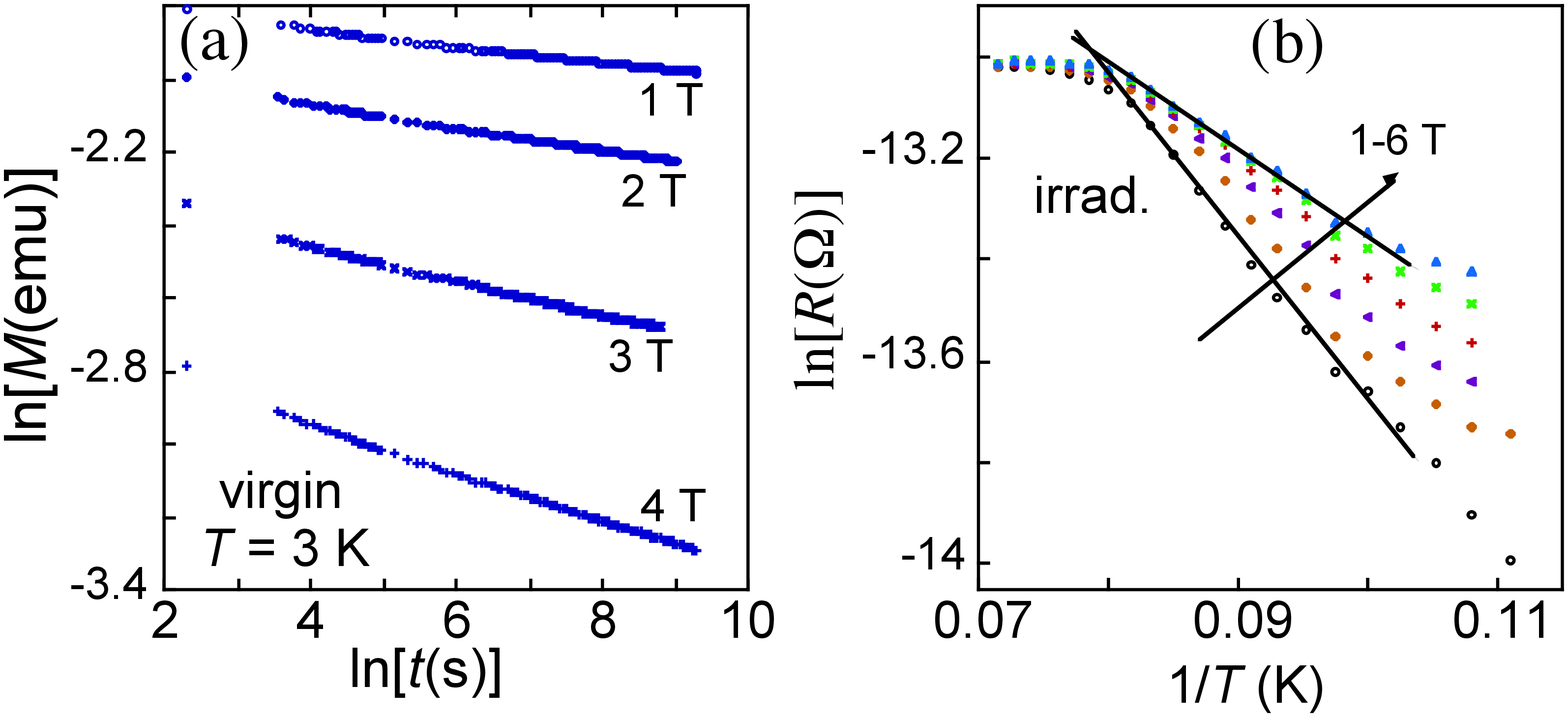}
\caption{a)Example magnetic relaxations curves as obtained for the virgin sample; b) Isofield magnetoresistivity curves as obtained for the irradiated sample.}
\label{fig3}
\end{figure}

Figure 4 shows a plot of the activation pinning energy $U_0$ as estimated from flux-creep data in the irreversible regime and from magnetoresistivity data in the reversible regime for both samples. The plots shown in Fig. 4 allow one to compare and to observe that values of $U_0$ obtained from flux-creep (in the irreversible regime) are quite 
different and lower than values obtained in the flux-flow regime. As well as the 
difference between the activated pinning energy within the different regimes, it is also clear that irradiation significantly reduces $U_0$. This result is rather interesting, 
motivating us to further explore the influence of irradiation on the pinning force and the pinning mechanism. 
 \begin{figure}[t]
\includegraphics[width=\linewidth]{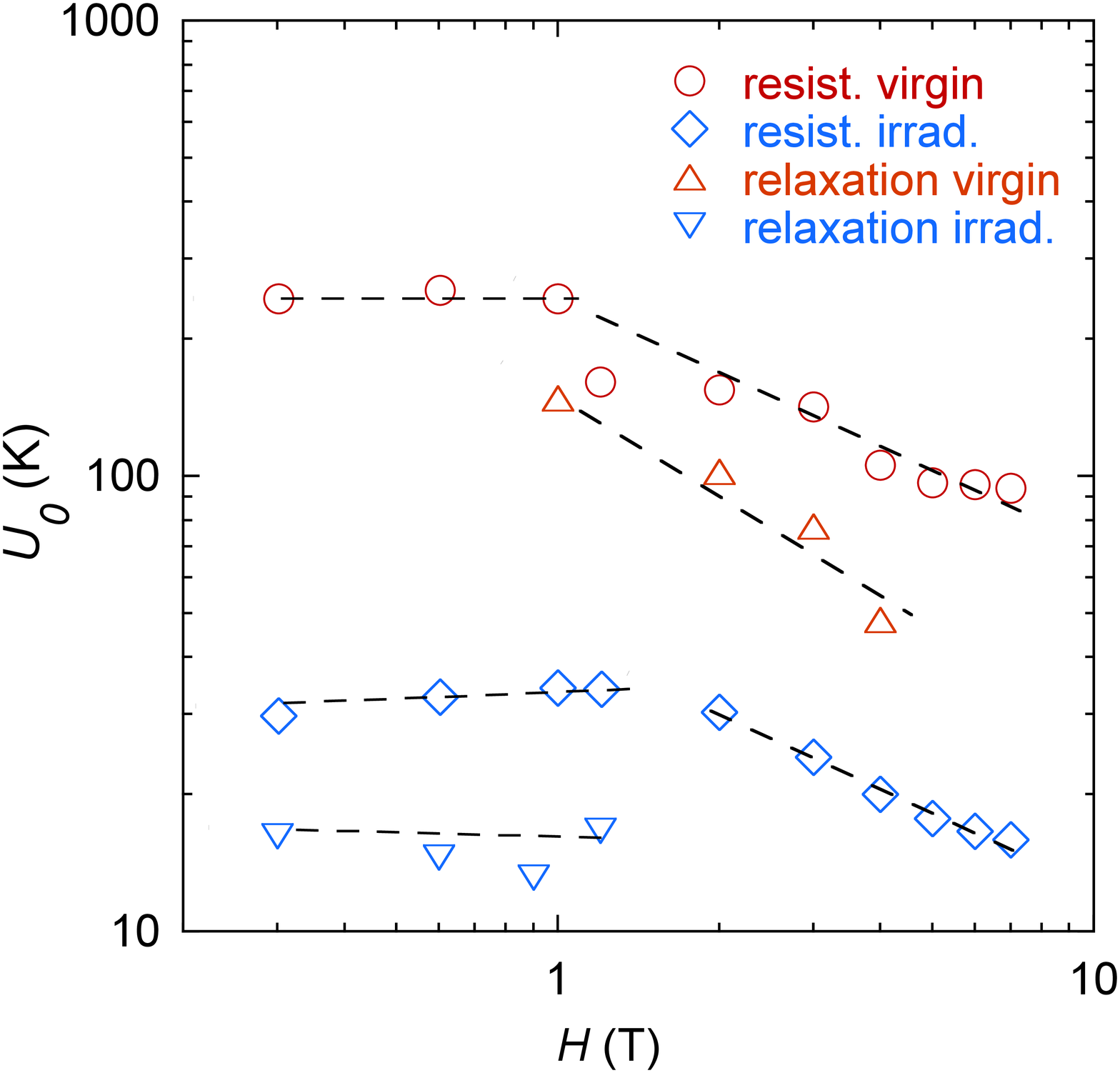}
\caption{$U_0$ as obtained from flux-flow and from flux-creep on both samples is plotted against the magnetic field. Dotted lines are guide-for-the eye only.}
\label{fig4}
\end{figure}

Figure 5 shows plots of several isothermic curves of the normalised volume pinning force, $Fp$/$Fp_{max}$, as a function of the  reduced field $h$ = $H$/$H_{irr}$ as obtained for both samples, where $Fp$ = $J_c$x$B$ and $B$ is the induction magnetic field (it is assumed that $B$ = $H$ which is a good approximation for fields far from the Meissner region). It should be noted that the many different curves in each figure appear to collapse to a single curve. Solid lines appearing on Fig. 5a (virgin crystal) and Fig. 5b (irradiated crystal) are obtained by fitting the data to the expression $Fp$ $\approx$ $h^p$ (1-$h$)$^q$ \cite{Dew} where $p$ and $q$ are fitting parameters which provide insight into the type of pinning. Moreover $p$/($p$+$q$) represents the field, $h_{max}$, for which the maximum value of  $Fp$/$Fp_{max}$ occurs. Results of the fittings are: $p$ =  0.5 and $q$ = 3.4 for the virgin crystal (Fig. 5a) and $p$ = 0.36 and $q$ = 2 for the irradiated sample (Fig. 5b). While the values of $p$ and $q$ for the virgin crystal are difficult to explained in terms of possible values expected for different types of pinning \cite{Dew}, the values of $p$ and $q$ for the irradiated sample suggest that point pinning dominates in this case. It is interesting to note that the value of $h_{max}$ virtually did not change with irradiation. 
 
It should be mentioned that the low value of $h_{max}$ for both samples, suggests that pinning is of $\delta _L$pin type (spatial variations in the mean free path of the carriers) for both samples. To check this finding, we plot in Fig. 6 values of  the so called remanent state $J_c$ for $H$ = 0 (normalised by $J_c$ at $T$ = 2 K), $J_c$/$J_c$(2K), as a function of the reduced temperature, $T$/$T_c$, which are obtained from the remanent magnetisation. It is interesting to observe that the normalised values of $J_c$ for both samples almost superpose each other, suggesting that despite the differences observed in $U_0$, $T_c$ and $J_c$, due to irradiation, the same type of pinning is dominant in both samples.
As performed in Ref. \cite{griessen} we plot in Fig. 6 the values of $J_c$/$J_c$(2K) calculated assuming that pinning is of $\delta_L$pin type for which the temperature dependence is given by (1-$t^2$)$^{2.5}$(1+$t^2$)$^{-0.5}$ (red dotted line) and assuming that pinning is of  $\delta_{T_c}$pin type (random variation of $T_c$ within the sample volume) which has a temperature dependence given by (1-$t^2$)$^{7/6}$(1+$t^2$)$^{5/6}$ (blue dashed line).
A comparison between the curves suggests that $\delta_L$pin is the predominant type of pinning for both samples, which agrees with the low value of $h_{max}$ observed for both samples in Fig. 5. 
\begin{figure}[t]
\includegraphics[width=\linewidth]{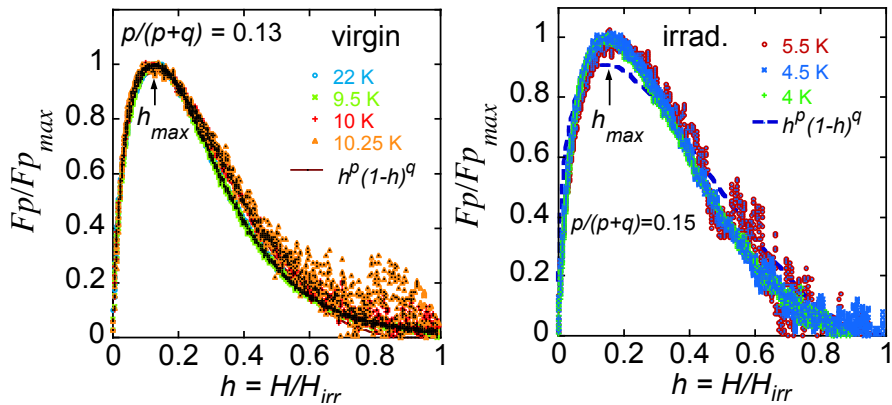}
\caption{The normalised pinning force is plotted against the reduced field $h$ for: a) virgin sample ; b) irradiated sample.}
\label{fig5}
\end{figure}

 \begin{figure}[t]
\includegraphics[width=\linewidth]{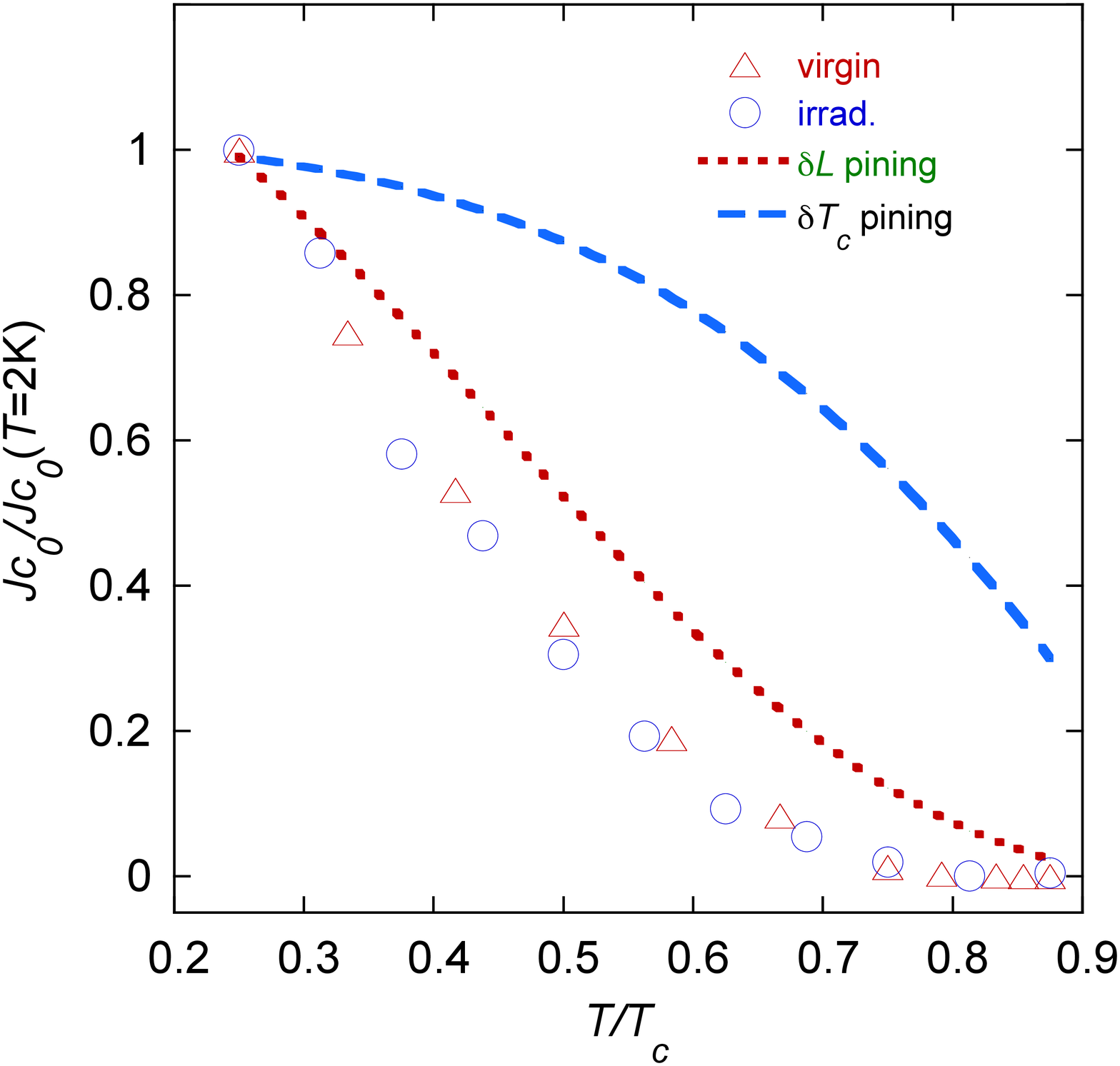}
\caption{The normalised remanent critical current is plotted against temperature for both samples. Dotted and dashed lines represents the values expected for $\delta_L$pin and $\delta_{T_c}$pin.}
\label{fig6}
\end{figure}

Finally we present in Fig. 7 a plot of the temperature dependence of the irreversibility line, IL, $H_{irr}$ vs $T$, for the studied samples, where values of $H_{irr}$ are obtained by extracting from each hysteresis curve the field for which magnetisation become reversible. A simple inspection by eye shows the effect of irradiation on the IL. Since irradiation is usually associated with an increase disorder which in the present case appears in conflict with the large reduction of $J_c$ as well $U_0$ after irradiation, it is interesting to fit the IL of Fig. 7 to an expression presented in ref. \cite{baruch}, which  was developed considering a parameter that measures the disorder ($n_p$). Solid lines in the curves of Fig. 6 represents the best fit of each IL to the expression,
$$1-t-b+2[n_p(1-t)^2 b/4\pi] [3/2-(4\pi t \sqrt{2Gi}/(n_p(1-t)^2))]=0$$ 
where $t$=$T$/$T_c$ is the reduced temperature, $b$=$H$/$H_{c2}(0)$ is the reduced field, $n_p$ is a parameter measuring the disorder and $G$i is a parameter associated to the Ginzburg number measuring the strength of thermal fluctuations. The best fit to the virgin curve is obtained assuming $T_c$ = 13 K producing the values $H_{c2}(0)$ = 250 kOe, $n_p$ = 0.005 and $Gi$ = 5x10$^{-5}$. For the irradiated curve, the best fit is obtained assuming $T_c$ = 10 K producing the values $H_{c2}(0)$ = 130 kOe, $n_p$ = 0.0008 and $Gi$ = 5x10$^{-5}$. Similar values of $n_p$, but a relatively lower value of $Gi$ were obtained
for Ba(Fe-Ni)$_2$As$_2$ samples \cite{jesus}. It should be noted that the fittings produced the same value of $Gi$ for both curves, but interestingly a much smaller value of the disorder parameter $n_p$ for the irradiated curve, which is consistent with the observed decrease in $J_c$ as well as in $U_0$ after irradiation. This model suggests that irradiation creates a high density of defects with very low activation pinning energy such that the overall effect is to reduce the average volume pinning and the effective disorder. 
\begin{figure}[t]
\includegraphics[width=\linewidth]{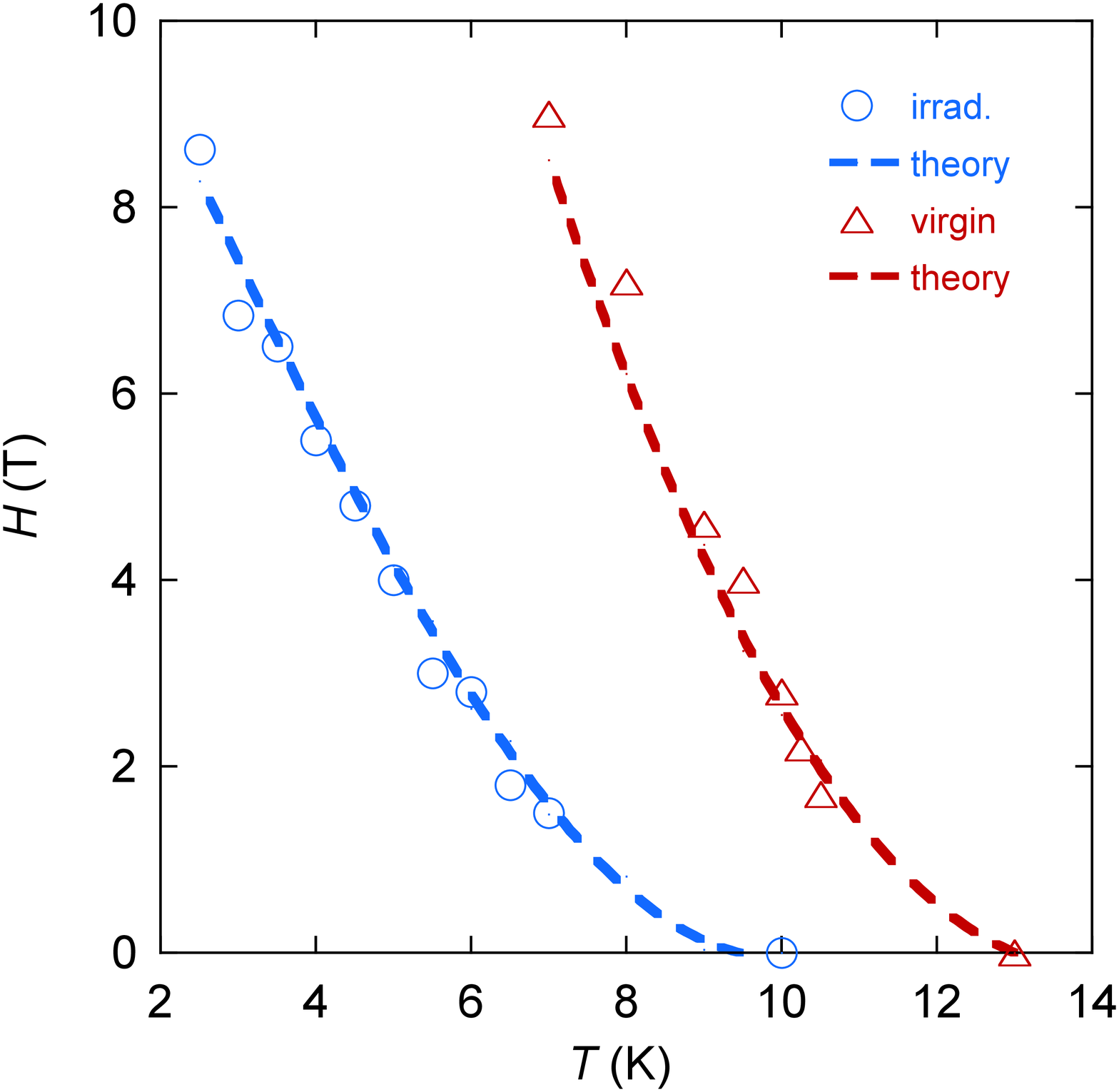}
\caption{$H_{irr}$ is plotted against the temperature for both samples. Dashed lines in each curve represents a fitting to the theory (see text for details).}
\label{fig7}
\end{figure}

\section{Conclusions}
In conclusion, our study shows that proton irradiation produced many changes in intrinsic parameters of the superconductor Ba(Fe$_{0.96}$Co$_{0.04}$)$_2$As$_2$. We find that irradiation in this case produces an apparent healing of the crystal in that the critical current density, the extracted $U_0$ and the irreversibility line are all reduced after irradiation. We observe that values of $U_0$ obtained from flux creep are lower than those obtained from the thermal assisted flux flow region. Despite these changes, analysis based on the pinning force and on the temperature dependence of $J_c$ suggest that the same type of pinning, $\delta_L$pin, remains after irradiation. One of the interesting observations is that the resulting defect character in the irradiated crystal causes the disappearance of the SMP, shedding light on the origin of this widely discussed feature. 
~

${\bf Acknowledgements}$
 
This work was supported by CAPES, Science without Borders program, grant number 88881.030498/2013-01. LC also acknowledges support from the EPSRC. LG, SSS and ADA were supported by FAPERJ and CNPq. SJS was supported by Fullbright. AS was supported by the U.S. Department of Energy, Office of Science, Basic Energy Sciences, Materials Science and Engineering Division.


\end{document}